\documentclass{PoS}

\title{Top physics at CLIC and ILC}

\ShortTitle{Top physics at CLIC and ILC}

\author{\speaker{Aleksander Filip \.Zarnecki} \hspace{\textwidth}
 on behalf of the CLICdp collaboration and the ILC Physics and Detector Study \\
        Faculty of Physics, University of Warsaw\\
        E-mail: \email{Filip.Zarnecki@fuw.edu.pl}}

\abstract{
  Measurements of top quark production at $e^+e^-$ colliders
  can provide a leap in precision in our knowledge of top quark properties
  and open a new window on physics beyond the Standard Model. In this
  contribution the top quark physics prospects of linear colliders is
  reviewed. Progress in detailed full-simulation studies is reported
  for the highlights of the program. We present the prospects for a
  measurement of the top quark mass and width in a scan of the beam
  energy through the pair production threshold, and discuss new
  studies of alternative measurements in continuum production, which
  are also capable of a precise determination of the mass in a
  rigorously defined mass scheme. A precision of ~50 MeV on the
  $\overline{MS}$ mass is expected when taking into account the dominant
  systematic uncertainties. Another key measurement is the study of
  the top quark couplings to electroweak gauge bosons, where form
  factors can be determined to 1\% precision, an order of magnitude
  better than those from the full LHC program. New results extend the
  prospects to different center-of-mass energies and to CP violating
  form factors. Finally, new studies are presented indicating the
  possibility to detect Flavour Changing Neutral Current decays
  of the top quark at linear colliders, such as the decay $t \rightarrow cH$,
  to a branching ratio of $BR(t\rightarrow cH) \sim 10^{-5}$.
}

\FullConference{38th International Conference on High Energy Physics\\
		3-10 August 2016\\
		Chicago, USA}

\begin{document}

\section{Introduction}

Top physics, together with Higgs boson studies and searches for Beyond
the Standard Model (BSM) phenomena, is one of the three pillars of the
research program for future high energy $e^{+} e^{-}$ colliders.
As top quark is the heaviest known elementary particle, with an
expected value of the Yukawa coupling of the order of one, the precise
determination of its properties is a key to the understanding of
electroweak symmetry breaking. 
Together with the Higgs mass, the top mass value is crucial for testing
the vacuum stability of the Standard Model.
Determination of top properties is also essential for many ``new
physics'' searches, as top quark gives large loop contributions to many
precision measurements sensitive to BSM effects.
Finally, as top is the only quark not forming a hadronic state, its
production and decays allow for detailed tests of QCD calculations.
Both future linear colliders, ILC and CLIC,
provide the opportunity to study the top quark with unprecedented
precision via direct production of $t\bar{t}$ pairs in  $e^{+} e^{-}$
collisions. 

\section{Colliders and Experiments}

The International Linear Collider (ILC) project is based on the 
technology of superconducting accelerating cavities.
In the Technical Design Report (TDR) completed in 2013 \cite{Adolphsen:2013kya},
construction of a machine with a centre of mass energy of 500 GeV and
a footprint of 31 km was proposed, with a possible upgrade to 1 TeV.
The baseline design includes polarisation for both $e^-$ and $e^+$ beams,
of 80\% and 30\%, respectively.
The running scenario H-20, which was selected as the most promising
\cite{Barklow:2015tja},  assumes collecting 500 fb$^{-1}$ of
data at 500 GeV in the initial ILC stage, followed by  200 fb$^{-1}$
collected at the top pair production threshold and 500  fb$^{-1}$ at
250 GeV in the first 8 years of running.
After the luminosity upgrade, an additional 3500 fb$^{-1}$ at 500 GeV and
1500 fb$^{-1}$ at 250 GeV  could be accumulated in about 11 years.

The Conceptual Design Report (CDR) for the Compact Linear Collider (CLIC)
was presented in 2012 \cite{Aicheler:2012bya}.
CLIC is based on the two-beam acceleration scheme, required to
generate a high RF gradient of about 100 MV/m. 
In the recently updated implementation plan for CLIC \cite{CLIC:2016zwp},
a construction in three stages is proposed, with 5 to 7 years of data
taking at each stage.
The first stage with a footprint of 11 km will allow to reach an energy
of  380 GeV, giving access to most Higgs boson and top quark measurements.
The plan assumes collecting 500~fb$^{-1}$ at 380 GeV with additional
100 fb$^{-1}$ collected at the $t\bar{t}$ threshold. 
The second and third construction stages at around 1.5 and 3 TeV,
with expected integrated luminosities of  1500 fb$^{-1}$ and  3000 fb$^{-1}$,
will focus on the searches for BSM phenomena. However, they will also
open possibilities for additional Higgs and top-quark measurements,
such as the direct determination of the top Yukawa coupling.
Polarisation of the electron beam is currently included in the CLIC
baseline design, while positron polarisation is considered as a
possible upgrade.

The detector concepts proposed for ILC and CLIC are based on 
jet reconstruction and jet energy measurements with the ``Particle
Flow'' approach \cite{Thomson:2009rp}. 
Single particle reconstruction and identification exploits high
calorimeter granularity, and
the best possible jet energy estimate is obtained by combining
calorimeter measurements for neutral particles with precise track momentum
measurements for the charged ones.
Very efficient flavour tagging is possible with a high precision pixel vertex
detector placed very close to the interaction point.
The background to processes with missing energy can be strongly suppressed
thanks to very good detector hermeticity, with instrumentation extending
down to a minimum angle of $\theta_{min} \sim 5$~mrad.
Although based on different technology choices, a similar performance
is expected for the two ILC detector concepts: ILD and SiD
\cite{Behnke:2013lya}.
Detailed simulation studies for CLIC were initially based on the
adopted ILC concepts~\cite{Linssen:2012hp}.
Recently, however, a dedicated detector model has become available
for the analysis.

\section{Top Quark measurements}

\subsection{Mass and width}

\begin{figure}[tb]
\begin{center}
  \includegraphics[width=0.45\textwidth]{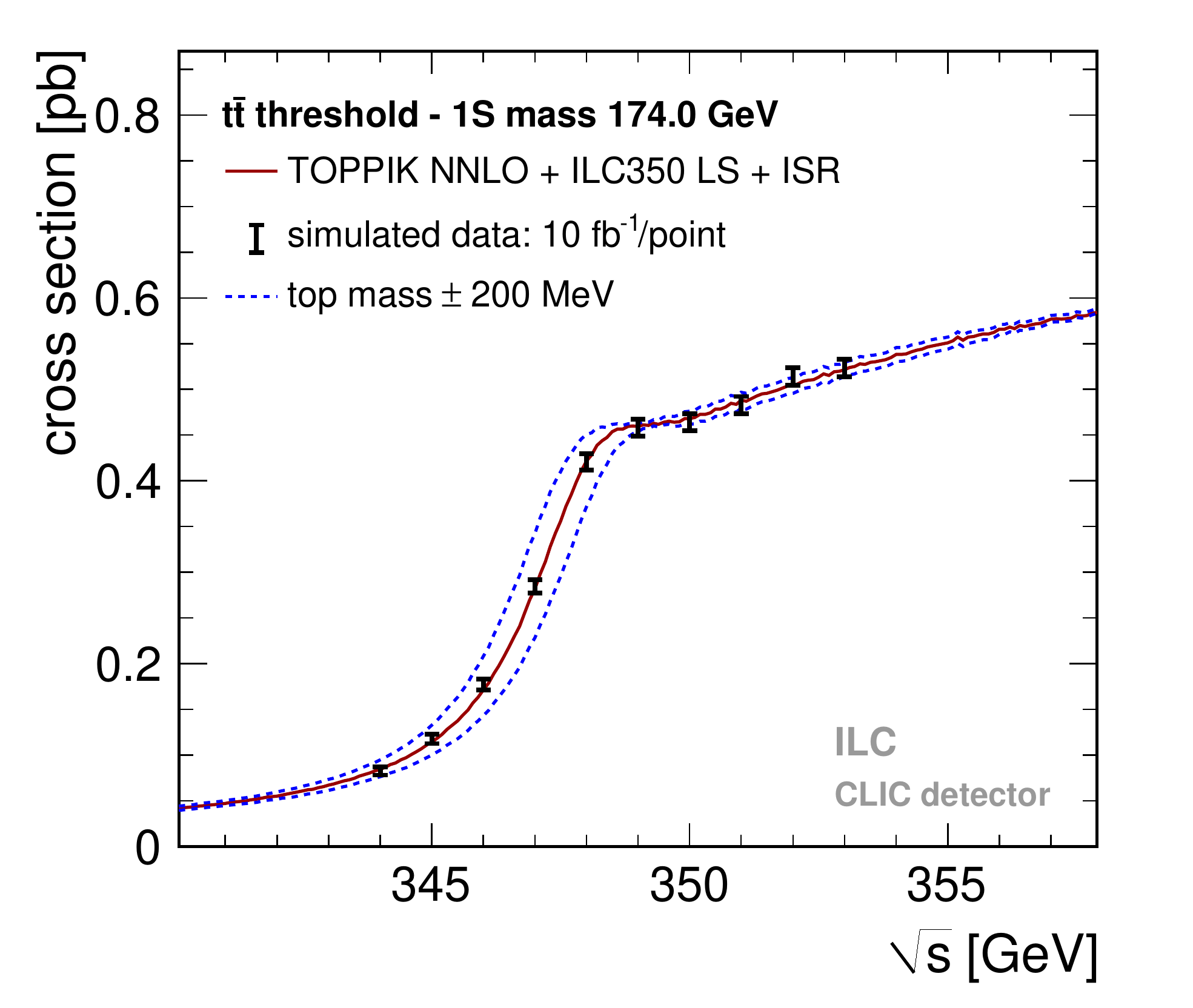}
  \includegraphics[width=0.45\textwidth]{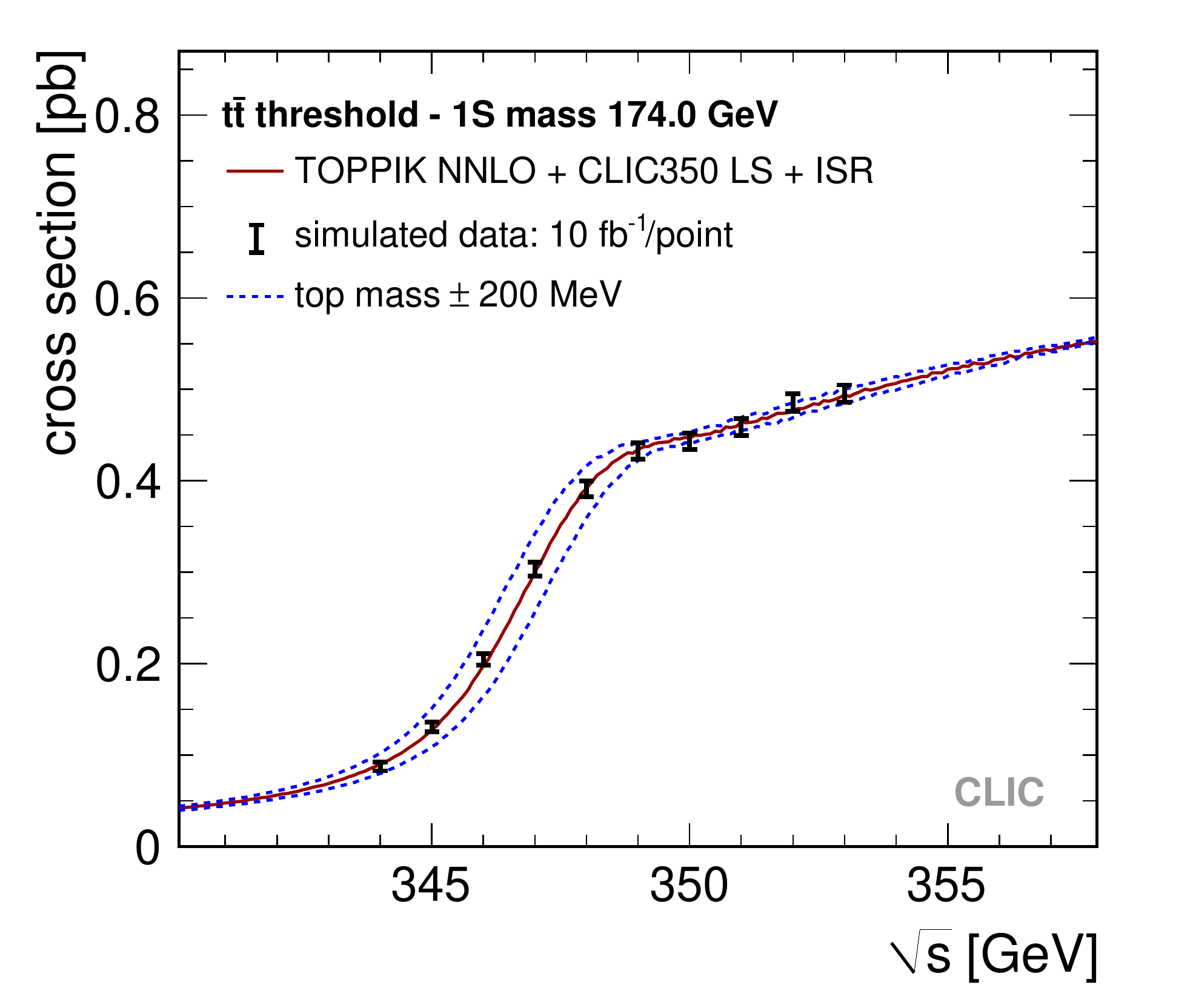}
\end{center}
\vspace{-0.7cm}
\caption{Simulated cross section measurements for top pair production
  at the threshold (background subtracted) with the luminosity spectra
  expected for ILC (left) and  CLIC (right). Threshold scan with 
  measurements at 10 energy points with 10 fb$^{-1}$ each, 
  simulated for the generator mass of 174 GeV (full line) as well as for
  a shift in mass of $\pm$200 MeV (dotted lines).
  Figure taken from \cite{Seidel:2013sqa}.
}
\label{fig-threshold}
\end{figure}

The dependence of the theoretical top pair production cross section on the
centre of mass energy shows a clear resonance-like structure at the threshold,
corresponding to a narrow $t \bar{t}$  bound state.
The shape of the cross section is very sensitive to top quark properties and
model parameters: mass $m_t$, width $\Gamma_t$,
Yukawa coupling $y_t$ and strong coupling $\alpha_s$.
In spite of the significant cross section smearing due to luminosity
spectra and initial state radiation (ISR), a precise $m_t$
measurement is possible already with 100 fb$^{-1}$. 
Detailed simulation studies were performed for both ILC and CLIC
\cite{Seidel:2013sqa,Horiguchi:2013wra} showing that with cross
section measurements at 10 different values of collision energy, each
with 10 fb$^{-1}$ of data (see Fig.~\ref{fig-threshold}), 
a statistical accuracy of 15--20 MeV can be obtained.
Theoretical uncertainties are currently estimated
to be at the level of 40 MeV \cite{Simon:2016htt,Simon:ichep2016},
uncertainty from $\alpha_s$  to about 30 MeV (for today's world average)
and experimental uncertainties (backgrounds, etc.) are estimated on
the 10--20 MeV level. 
The total uncertainty expected on $m_t$ is $\sim$50 MeV,
while $\Gamma_t$ could be extracted to about
40~MeV~\cite{Ishikawa:toplc2015}.

The main advantage of $m_t$ determination from the threshold
scan is that the extracted parameter is well defined from the theoretical
point of view. 
Direct reconstruction of the top quark mass from its decay products has been
considered for energies above the threshold
(continuum)~\cite{Seidel:2013sqa}.
Competitive statistical precision can be expected ($\sim$80 MeV
for 100~fb$^{-1}$ at 500~GeV).
However, the measurement suffers from significant theoretical
uncertainties when converting the extracted $m_t$ value to a particular mass
scheme (as for the ``standard'' measurements at LHC). 

Therefore, other methods of the top quark mass determination are also
being studied. 
At high energies, the top quark mass could be reconstructed from 
radiative events $e^+ e^- \rightarrow t \bar{t} \gamma $. Measurement of
a threshold in the ISR photon energy distribution,
corresponding to the $t \bar{t}$ production threshold, should
allow $m_t$ extraction with a statistical precision of the order of
100 MeV \cite{Vos:2016til}.
Other methods proposed recently are based on the reconstruction of the
$b$-jet energy distribution (``one prong'' method)~\cite{Agashe:2016bok}
or on the event shape analysis~\cite{Butenschoen:2016lpz}.

\subsection{Electroweak couplings}

\begin{figure}[tb]
\begin{center}
  \includegraphics[width=0.44\textwidth]{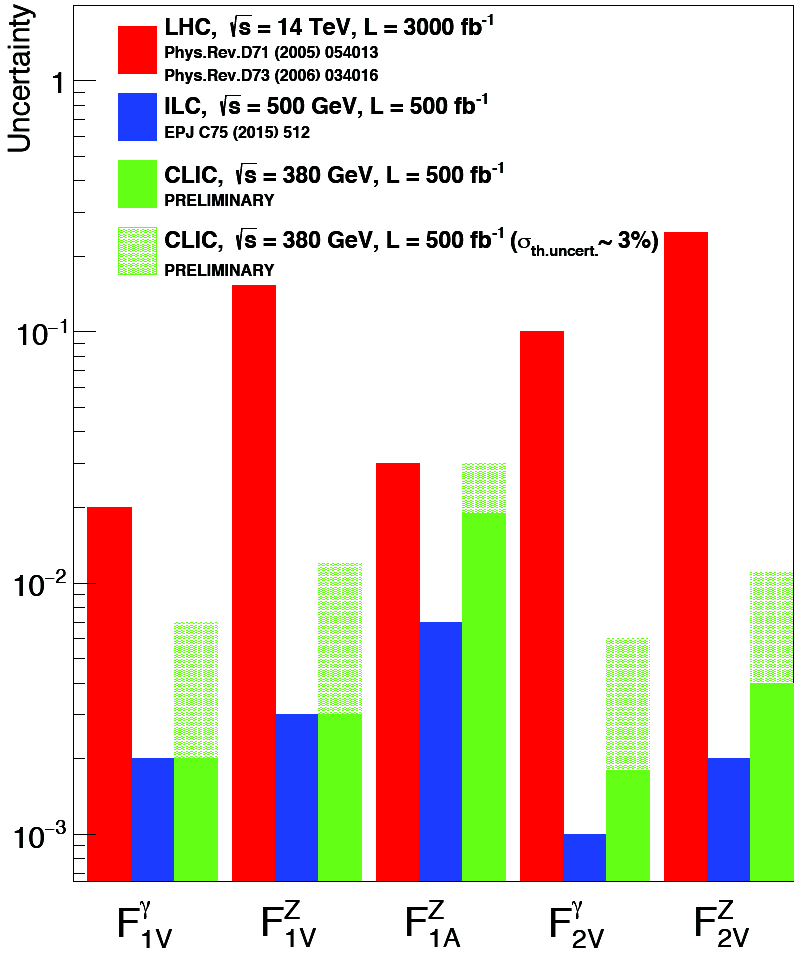}
  \includegraphics[width=0.44\textwidth]{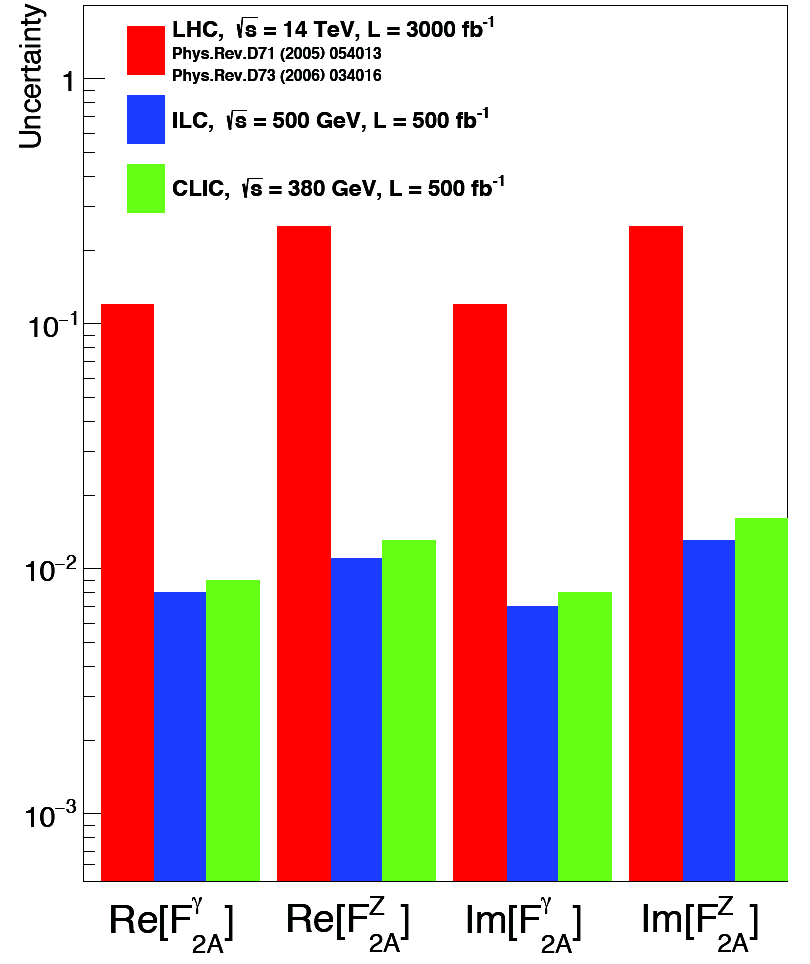}
\end{center}
\vspace{-0.7cm}
\caption{ Comparison of the uncertainties on the measured top-quark
  form factors (assuming SM values for the remaining form factors)
  expected for HL-LHC, ILC and CLIC. Considered are parity conserving
  (left) and parity violating (right) couplings. The form factors are
  extracted from the measured cross sections, forward backward
  asymmetry and  helicity angle distribution in top quark decays.
}
\label{fig-couplings}
\end{figure}

The measurement of the top pair production above threshold is
sensitive to the top quark electroweak couplings but also to possible higher
order corrections due to different ``new physics'' scenarios.
To constrain BSM contributions we consider the general form of
the top quark couplings to $Z$ and $\gamma$, which can be written in
terms of eight form factors (only three of them contributing to the
$t\bar{t}$ production in the Standard Model).
These form factors can be constrained through measurements of the
total $t\bar{t}$ production cross-section, forward-backward asymmetry and
helicity angle distribution in top quark decays, 
for two polarisation combinations: $e^-_L e^+_R$ and $e^-_R e^+_L$.
Results of a detailed simulation study performed for ILC
\cite{Amjad:2015mma} and extended to CLIC \cite{Perello:ecfa2016} 
show that already with 500 fb$^{-1}$, all form factors can be
constrained to 1\%  or below, while the HL-LHC will only be
able to set limits in the 10\% range (see Fig.~\ref{fig-couplings}).
Further improvement of the linear collider precision is possible by improving 
$b$-jet charge reconstruction and particle identification, so that
fully hadronic $t\bar{t}$ decays can also be included in the analysis
\cite{bilokin:toplc2016}.
With 20 years of data taking at the ILC, the statistical precision of
the coupling determination could be reduced to the level of 0.2\%.
However, to profit from an ILC luminosity upgrade 
theoretical and experimental uncertainties have to be controlled to
the per mille level. 

\subsection{Yukawa coupling}

The top Yukawa coupling can already be constrained from the scan of the
top pair production threshold, as the cross section includes a
contribution of about 9\% from the virtual Higgs exchange
\cite{Horiguchi:2013wra}. 
With 100~fb$^{-1}$ of data the coupling can be extracted with a statistical
uncertainty of about 6\%, assuming the $\alpha_s$ value can be constrained
from other measurements.
The precision on $y_t$ will therefore be dominated by theoretical
uncertainties which are currently estimated to be at the level of 20\%.
When running at higher energies (at or above 500 GeV),  $y_t$ 
can be directly extracted from the measurement of
$e^+e^- \rightarrow t\bar{t}H$ events.
Even with the excellent performance of linear collider detectors, which is
required for good background suppression and efficient event selection,
the measurement will be limited by statistics.
With 2$\times$500 fb$^{-1}$ of data collected at 500 GeV,
the ILD experiment should measure $y_t$ with a statistical uncertainty
of about 11\%,
which can be reduced to 6.4\% with 4000 fb$^{-1}$ \cite{sudo:toplc2016}.    
Significant improvement is expected when going to higher energies.
Already by running at 540 GeV, the statistical uncertainty can be reduced
by a factor of 2.
A precision of 4-5\% is also expected for ILC running at
1~TeV \cite{Price:2014oca} or CLIC running at 1.4~TeV.

\subsection{Rare decays}

Flavor-Changing Neutral Current (FCNC) top quark decays are
strongly suppressed in the Standard Model, with the expected
branching ratios
$BR(t \rightarrow  c\; X )   \sim   10^{-15}$ to $10^{-12}$
($X = \gamma, \; g, \; Z, \; H$).
Observation of any such decay would be a direct signature
for ``new physics''.
The decay $t\! \rightarrow\!  c\; H$ seems to be the most promising
channel, as an enhancement up to  $10^{-2}$ is possible~\cite{Agashe:2013hma}.
This decay is difficult to constrain at LHC, with expected HL-LHC
limits of  $BR < 2 \cdot 10^{-4}$ \cite{Agashe:2013hma,Atlas:FCNC}.
At  $e^+ e^-$ colliders top pair production events with FCNC decay
$t\! \rightarrow\!  c\; H$ can be identified based on the kinematic 
constrains and flavour tagging information.
Although the signal selection efficiency is limited by large overlap
in kinematic space with standard top pair events, 
parton level simulation results indicate that  with high integrated
luminosity this decay can be probed down to
$BR \sim 10^{-5}$~\cite{Vos:2016til}. 
A full simulation study is ongoing.

\section{Conclusions}

A precise determination of top quark parameters is crucial for the
validation of the Standard Model or any alternative BSM theory.
Linear $e^+ e^-$ colliders, ILC and CLIC,
offer unique opportunities for these measurements.
A scan of the top pair production threshold will allow for mass and
width measurements,
while direct extraction of Yukawa and electroweak couplings require running
at higher beam energies.
High precision and background suppression capabilities of ILC and CLIC
detectors will allow for per mille level measurements and searches for
rare processes. 
However, even in a clean environment of $e^+ e^-$ collisions,
top events reconstruction remains very challenging and imposes stringent
requirements on the detector performance.

\end{document}